# Competition of Two Terms in Exchange Hamiltonian $H_{ex}=-2A_1\sum_{i<j}\vec{s}_i\cdot\vec{s}_j-2A_2\sum_{i<j}\vec{s}_i\cdot\vec{s}_j$ ($A_1>0$, $A_2<0$) for magnetic substances

WANG Yong-zhong

Institute of Metal Research, The Chinese Academy of Sciences, Shenyang 110016, People's Republic of China, E-mail address: yzhwang@imr.ac.cn

The unjustifiable or wrong in the previous magnetism theories has been indicated in this paper. For a N electrons system with Heisenberg exchange integral $A = A_1 + A_2$ ($A_1>0, A_2<0$), the correct exchange Hamilton should be $H_{ex}=-2A_1\sum_{i<j}\vec{s}_i\cdot\vec{s}_j-2A_2\sum_{i<j}\vec{s}_i\cdot\vec{s}_j$, not $H_{ex}=-2A\sum_{i<j}\vec{s}_i\cdot\vec{s}_j$ as in the previous magnetism theories. The role of the minor term in the exchange Hamilton was considered. Based on the principle of superposition of state, the eigenstate of the system with Heisenberg exchange integral $A = A_1 + A_2$ ($A_1>0, A_2<0$)

$$|X\rangle = \frac{1}{\sqrt{A_1^2+A_2^2}}(A_1|1\rangle+|A_2\|2\rangle)$$

and the energy (relative to exchange interaction) eigenvalue $E=-Nz(A_1-|A_2|)-\frac{2Nz|A_2|}{A_1+|A_2|}=-Nz(|A_2|-A_1)-\frac{2NzA_1}{A_1+|A_2|}$ were attained, where $z$ is the number of the nearest neighbours electrons, $|1\rangle$ means the state of the system when the spins of all electrons in the system arrange parallelly, $|2\rangle$ means the state when the spins of all electrons or the nearest neighbor electrons in the system arrange antiparallelly. When $A_1 = |A_2| \neq 0$, $|X\rangle = \frac{1}{\sqrt{2}}(|1\rangle+|2\rangle)$ and $E$ = -$NzA_1$, the system is in the spin glass (SG) state, the probabilities of parallel and antiparallel arrange for every pair of spins of electron of nearest neighbours in the system are equal. When $A_1 \neq |A_2|$, the probabilities are not equal, and there coexist the states FM and SG or AFM and SG,

$$|X\rangle = \frac{1}{\sqrt{A_1^2 + A_2^2}}[(A_1 - |A_2|)|1\rangle + |A_2|(|1\rangle + |2\rangle)] \text{ or}$$

$$|X\rangle = \frac{1}{\sqrt{A_1^2 + A_2^2}}[(|A_2| - A_1)|2\rangle + A_1(|1\rangle + |2\rangle)].$$ When FM and SG or AFM and

SG coexist, the energy of the system is lower than that when only FM or AFM exists as in previous theory. Weiss ferromagnetic state or Neel anti ferromagnetic state is just a special state when $A_1=0$ or $A_2=0$.

I. INTRODUCTION

All theories about the magnetism of substances, up to date, have met the following serious difficulties [1-4]: 1) Most of ferromagnetic substances have two magnetic transition temperatures, the paramagnetic Curie temperature and the ferromagnetic Curie temperature. The paramagnetic Curie temperature can be found experimentally according to the Curie-Weiss law and the ferromagnetic Curie temperature can be measured experimentally. The paramagnetic Curie temperature of a substance is higher than the ferromagnetic Curie temperature. All the theories could give the ferromagnetic Curie temperature $T_c$ determined by the overall exchange energy $A = A_1 + A_2$ ($A_1>0$, $A_2<0$), which unfortunately corresponds to the paramagnetic Curie temperature $\theta_p$ experimentally measured. It means that theory can not give correct ferromagnetic Curie temperatures of the magnetic substances. This is a big contradiction within the frame of these theories. 2) It was found that the temperature dependence of the reciprocal magnetic susceptibility (i.e., $1/\chi \sim T$ curve) of ferromagnetic materials shows an upward winding near Curie temperature [5,6]. It has been ascribed to the existence of the short - range order. But in practice, the short - range order leads to that the matter is magnetized more easily and thus must result in a downward winding of the $1/\chi \sim T$ curve near the Curie temperature. There must be another reason for this upward winding. The disorder freezing of the spins in the systems of spin glasses has been challenging to all theories in modern physics. It is also hard to find a good theoretical model to explain the re-entrant phenomenon in the spin glasses systems, such as $Au_{1-x}Fe_x$ [4].

Heisenberg exchange Hamiltonian[7~9] for a two electron system can be denoted by $H$ (Heisenberg) = $-2J_H \vec{s}_1 \cdot \vec{s}_2$, $J_H = \left\langle ab \left| \frac{e^2}{r_{12}} \right| ba \right\rangle - 2(S_{ab} + \Gamma_{ab})\langle b|V|a\rangle$, Heisenberg – Dirac exchange Hamiltonian[8] for N localizes spins can denoted by

$H_{H-D} = -\sum_{l \neq m}(J_{lm} - \frac{2|V_{lm}|}{U})\vec{s}_l \cdot \vec{s}_m$. The first terms of the two formulas which are positive definite favours ferromagnetic coupling, the second terms which are always negative definite favours the antiferromagnetic state. For convenience, in this paper we denote the first term and the second term as $A_1$ (>0) and $A_2$ (<0), respectively.

The spontaneous ordering of the spins or the atomic moments originated from the dominant term and minor term in the exchange integrals. But the role of the minor term in the exchange integrals for magnetic matters was not considered in the previous theories of magnetism. The role of the minor term in the exchange integrals was studied in ref. 10. The independent physical role of the two exchange energy $A_1$ and $A_2$ was considered, a unique phenomenological theory for ferromagnetism, antiferromagnetism and spin glass was developed. In this phenomenological theory the magnetic behaviors of matters was determined by the competition among the thermal motion, $A_1$ and $A_2$, such as two Curie temperatures in ferromagnetic materials, the freezing of the spin glasses and the re-entrant phenomenon in the spin glasses systems observed in experiments were given. When $A_1 = |A_2|$, $A = 0$, the matters is in the spin glass state; when $A_1 > |A_2|$, there is the coexistence of FM and SG; When $|A_2| > A_1$, there is the coexistence of AFM and SG. Therefore only considering the overall interaction of first term and second term in Heisenberg exchange Hamiltonian is unjustifiable, it is just the reason for the inconsistent of the previous theories with experiments. This object will be studied continuously from the first principle in this paper. In order to consider simultaneously the physical function of the first term and the second term, in this paper we propose a new exchange Hamiltonian involving the first term and the second term instead of the old exchange Hamiltonian only involving the overall interaction.

The remainder of this paper is arranged as follows. The Heisenberg exchange Hamiltonian and quantum state of the electrons system with exchange integral $A = A_1 + A_2$ ($A_1 > 0$, $A_2 < 0$) will be described in Sec. II and III. The Energy of the electrons system will be calculated in Sec. IV. Section V is the summry.

II  Heisenberg exchange Hamiltonian of the system with exchange

   integral  $A = A_1 + A_2$ ( $A_1 > 0$, $A_2 < 0$)

Heisenberg exchange Hamiltonian[7~9] for a two electron system can be denoted by $H$ (Heisenberg) = $-2J_H \vec{s}_1 \cdot \vec{s}_2$, where

$$J_H = \left\langle ab \left| \frac{e^2}{r_{12}} \right| ba \right\rangle - 2(S_{ab} + \Gamma_{ab})\langle b|V|a \rangle, \quad (1)$$

$$\left\langle ab\left|\frac{e^2}{r_{12}}\right|ba\right\rangle \equiv \int \phi_a^*(r_1)\phi_b^*(r_2)(\frac{e^2}{r_{12}})\phi_b(r_1)\phi_a(r_2)d^3r_1 d^3r_2 \ , \quad S_{ab} = \left\langle \phi_a(r)|\phi_b(r)\right\rangle = \int \phi_a^*(r)\phi_b(r)d^3r \ ,$$

$\Gamma_{ab} = \dfrac{\langle a|V|b\rangle}{\Delta E(a \to b)}$, $\langle b|V|a\rangle \equiv \int \phi_b^*(r)V(r)\phi_a(r)d^3r$, $V = \dfrac{Ze^2}{r_{1b}}$, $\phi_a(r)$ and $\phi_b(r)$ are the atomic orbital functions which are solutions of the corresponding Schrödinger equations, i.e. $H_a\phi_a(r_1) = E_a\phi_a(r_1)$, $H_b\phi_b(r_2) = E_b\phi_b(r_2)$, $H_a = \dfrac{p_1^2}{2m} - \dfrac{Ze^2}{r_{1a}}$, $H_b = \dfrac{p_2^2}{2m} - \dfrac{Ze^2}{r_{2b}}$, $E_a$ and $E_b$ are the energy eigenvalues.

Heisenberg – Dirac exchange Hamiltonian[8] for N localizes spins can denoted by

$$H_{H-D} = -\sum_{l \neq m}(J_{lm} - \frac{2|V_{lm}|}{U})\vec{s}_l \cdot \vec{s}_m \qquad (2)$$

where $J_{lm} = \left\langle \phi_l(r_1)\phi_m(r_2)\left|\dfrac{e^2}{r_{12}}\right|\phi_m(r_1)\phi_l(r_2)\right\rangle$,

$V_{lm} = \left\langle \phi_l\left|\dfrac{p^2}{2m} + V(r)\right|\phi_m\right\rangle$, $U = U_{mn}^{jl} = \left\langle \phi_j(r_1)\phi_l(r_2)\left|\dfrac{e^2}{r_{12}}\right|\phi_m(r_1)\phi_n(r_2)\right\rangle$.

The exchange integral $\left\langle ab\left|\dfrac{e^2}{r_{12}}\right|ba\right\rangle$ in eqn (1) which is always positive definite favours the ferromagnetic (triplet) state, $2(S_{ab} + \Gamma_{ab})\langle b|V|a\rangle$ which is always negative definite favours the antiferromagnetic state; the potential exchange between spins of two different atoms (first term of eqn (2)) which is positive definite favours ferromagnetic coupling, the kinetic exchange term (second term of eqn (2)) which is is always negative definite favours the antiferromagnetic state. For convenience, in this paper we denote the first term and the second term of eqn (1) or (2) as $A_1$ (>0) and $A_2$ (<0), respectively.

In the previous magnetism theories,magnetism of substances has been studied only based on the overall exchange Hamiltonian $H_{ex} = -2A\sum_{i<j}\vec{s}_i \cdot \vec{s}_j$, **where** $A = (\left\langle ab\left|\dfrac{e^2}{r_{12}}\right|ba\right\rangle - 2(S_{ab} + \Gamma_{ab})\langle b|V|a\rangle)$ in eqn (1) or $(\sum_{l\neq m} J_{lm} - \sum_{l\neq m}\dfrac{2|V_{lm}|}{U})$ in eqn (2), but this point of view is unjustifiable! Because the first term of eqn (1) and (2) favour the ferromagnetic coupling, the second term of eqn (1) and (2) favour the antiferromagnetic coupling, the first and the second terms have independent and different physical function respectively, and there is the competition of the two terms. Only considering overall interaction is equivalent to only considering the first term or the second term of eqn (1) and (2). If the overall interaction is positive, only considering overall interaction is equivalent to only considering the function of the first term of eqn (1) and (2); If the overall interaction is negative, only considering overall interaction is equivalent to only considering the function of the second term of eqn (1) and (2). So only considering the overall interaction of first term and second term is not justifiable, it is just the reason for the inconsistent of the previous theories with experiments. In order to consider simultaneously the physical function of the first term and the second term, we propose a new exchange Hamiltonian involving the first term and the second term instead of the old exchange Hamiltonian of the overall interaction. The new exchange Hamiltonian can be expressed as

$$H_{ex} = -2A_1 \sum_{i<j} \vec{s}_i \cdot \vec{s}_j - 2A_2 \sum_{i<j} \vec{s}_i \cdot \vec{s}_j$$

III. Quantum state of the electrons system with exchange integral $A = A_1 + A_2$ ( $A_1>0$, $A_2<0$ )

For the electrons system with Heisenberg exchange integral $A = A_1 + A_2$ ($A_1>0$, $A_2<0$), the exchange Hamilton

$$H_{ex} = -2A \sum_{i<j} \vec{s}_i \cdot \vec{s}_j \qquad (3)$$

which has been used in previous magnetism theories and

$$H_{ex} = -2A_1 \sum_{i<j} \vec{s}_i \cdot \vec{s}_j - 2A_2 \sum_{i<j} \vec{s}_i \cdot \vec{s}_j \qquad (4)$$

have completely different physical meaning. When $A>0$, Eq. (3) is just equivalent to the first term in Eq. (4), and when $A<0$, it is just equivalent to the second term in Eq. 4). Therefore the exchange Hamilton $-2A_1 \sum_{i<j} \vec{s}_i \cdot \vec{s}_j - 2A_2 \sum_{i<j} \vec{s}_i \cdot \vec{s}_j$ is only correct for this system and it will be used in this paper.

When the spins of all electrons in the system arrange parallelly, the state of the system is expressed with Dirac symbol $|1\rangle$ [11], the magnetic moment M of system is an observable, the value is $M_1$; when the spins of all electrons arrange antiparallelly, the state of the system is expressed with Dirac symbol $|2\rangle$, M is also an observable, the value is $M_2$. So $|1\rangle$ and $|2\rangle$ are the eigenstates belonging to the different eigenvalues $M_1$ and $M_2$ for the same dynamical variable M. According to Dirac[11], $|1\rangle$ and $|2\rangle$ are orthogonal each other. For a system of two electrons[12], $|1\rangle = \frac{1}{\sqrt{2}}[|\uparrow\rangle_a|\uparrow\rangle_b \pm |\downarrow\rangle_a|\downarrow\rangle_b]$, $|2\rangle = \frac{1}{\sqrt{2}}[|\uparrow\rangle_a|\downarrow\rangle_b \pm |\downarrow\rangle_a|\uparrow\rangle_b]$, and they constitute a complete set of eigenstates.

If The exchange Hamilton $H_{ex} = -2A \sum_{i<j} \vec{s}_i \cdot \vec{s}_j$ was used, then when $A_1>0$ $H_{ex}|1\rangle = -NzA_1|1\rangle$, $A_2<0$ $H_{ex}|2\rangle = -Nz|A_2||2\rangle$. Now the new exchange Hamilton $H_{ex}$ of the system consists of $H_{ex1}$ and $H_{ex2}$, where

$H_{ex1} = -2A_1 \sum_{i<j} \vec{s}_i \cdot \vec{s}_j$ and $H_{ex2} = -2A_2 \sum_{i<j} \vec{s}_i \cdot \vec{s}_j$. Because $A_1>0$ and $A_2<0$, $H_{ex1}|1\rangle = -NzA_1|1\rangle$ and $H_{ex2}|2\rangle = -Nz|A_2||2\rangle$. When $A_1 \neq 0$ and $A_2 \neq 0$, there exists the probability of parallel arrangement and the probability of antiparallel arrangement for every pair of spins of the nearest neighbors in the system, and the value of the probability is dependent on the ratio between $A_1$ and $|A_2|$. So every pair of spins is partly in the quantum state of $|1\rangle$ and partly in the $|2\rangle$ and there only exist the two situations (similar to the situation of a photon passing through a crystal of tourmaline, which polarized obliquely to the the optic axis). Therefore the eigenstate of the system is neither $|1\rangle$ nor $|2\rangle$, but a linearity superposition of them according to the general principle of quantum mechanics. It can be represented as $|X\rangle = a|1\rangle + b|2\rangle$, where a/b= $A_1/|A_2|$。 After normalizing we have

$$|X\rangle = \frac{1}{\sqrt{A_1^2 + A_2^2}} (A_1|1\rangle + |A_2||2\rangle) \qquad (5)$$

IV.　Energy of the electrons system with exchange integral $A = A_1 + A_2$

The eigenequation is

$$H_{ex} \frac{1}{\sqrt{A_1^2 + A_2^2}} (A_1|1\rangle + |A_2||2\rangle) = E \frac{1}{\sqrt{A_1^2 + A_2^2}} (A_1|1\rangle + |A_2||2\rangle) \qquad (6)$$

or

$$(H_{ex1} + H_{ex2})(A_1|1\rangle + |A_2||2\rangle) = E(A_1|1\rangle + |A_2||2\rangle) \qquad (7)$$

where $E$ is the eigenvalue of $H_{ex}$, i.e. the energy relative to the exchange energy for the system, $H_{ex1} = -2A_1 \sum_{i<j} \vec{s}_i \cdot \vec{s}_j$ and $H_{ex2} = -2A_2 \sum_{i<j} \vec{s}_i \cdot \vec{s}_j$. The left terms of equal-sign in Eq. (7) give

$$H_{ex1}(A_1|1\rangle + |A_2||2\rangle) + H_{ex2}(A_1|1\rangle + |A_2||2\rangle) =$$
$$-Nz A_1^2 |1\rangle - Nz|A_2|^2|2\rangle + |A_2|H_{ex1}|2\rangle + A_1 H_{ex2}|1\rangle \qquad (8)$$

Applying $H_{ex1}$ to $|2\rangle$ turns the direction of $|2\rangle$ towards the direction of $|1\rangle$ for an angle $\beta$ [11], and applying $H_{ex2}$ to $|1\rangle$ turns the direction of $|1\rangle$ towards the direction of $|2\rangle$ for an angle $\alpha$. Then the $H_{ex1}|2\rangle$ and $H_{ex2}|1\rangle$ can be represented as a linearity superposition of $|1\rangle$ and $|2\rangle$

$$H_{ex1}|2\rangle = -Nz A_1 (\sin\beta |1\rangle + \cos\beta |2\rangle) \qquad (9)$$

$$H_{ex2}|1\rangle = Nz|A_2|\ (\sin\alpha\ |2\rangle + \cos\alpha\ |1\rangle\ ) \qquad (10)$$

The ratio between the absolute values of the weights of $|1\rangle$ in Eq. (9) and the weights of $|2\rangle$ in Eq. (10) must be equal to $A_1/|A_2|$, i.e. the ability of turning state vector $|2\rangle$ or $|1\rangle$ for $H_{ex2}/A_1$ or $H_{ex1}/|A_2|$ is the same, so $\sin\beta = \sin\alpha$. Upon substitution of Eq. (9) and Eq. (10) in Eq.(8), we have

$$(H_{ex1}+H_{ex2})\ (A_1|1\rangle+|A_2||2\rangle)$$
$$= (-A_1-|A_2|\ \sin\alpha+|A_2|\ \cos\alpha)\ A_1\ |1\rangle + (-|A_2|-A_1\cos\alpha +A_1\sin\alpha)\ |A_2|\ |2\rangle$$

From Eq.(7) we attain

$$E = -NzA_1 + Nz|A_2|\ (\cos\alpha - \sin\alpha)$$
$$= -Nz|A_2| - NzA_1\ (\cos\alpha - \sin\alpha) \qquad (11)$$

And formula (11) gives

$$\cos\alpha - \sin\alpha = \frac{A_1-|A_2|}{A_1+|A_2|} = k \qquad (12)$$

and

$$E = -NzA_1 + Nz|A_2|\frac{A_1-|A_2|}{A_1+|A_2|} = -Nz|A_2| + NzA_1\frac{|A_2|-A_1}{A_1+|A_2|} \qquad (13)$$

$\cos\alpha = \frac{k}{2} \pm \frac{1}{2}\sqrt{2-k^2}$ and $\sin\alpha = \pm\frac{1}{2}\sqrt{2-k^2} - \frac{k}{2}$ can be obtained from Eq. (11).

The Eq.(13) and Eq.(5) can be rewritten as

$$E = -Nz|A_1+A_2| - \frac{2Nz|A_2|}{A_1+|A_2|} = -Nz|A_1+A_2| - \frac{2NzA_1}{A_1+|A_2|} \qquad (12)$$

and

$$|X\rangle = \frac{1}{\sqrt{A_1^2+A_2^2}}[(A_1-|A_2|)|1\rangle + |A_2|(|1\rangle+|2\rangle)] \qquad \text{for } A_1>A_2 \qquad (13)$$

or

$$|X\rangle = \frac{1}{\sqrt{A_1^2+A_2^2}}[(|A_2|-A_1)|2\rangle + A_1(|1\rangle+|2\rangle)] \qquad \text{for } A_1<-A_2 \qquad (14)$$

It can be seen from Formula (12) that The energy eigenvalue obtained from the exchange Hamilton $-2A_1\sum_{i<j}\vec{s}_i\cdot\vec{s}_j - 2A_2\sum_{i<j}\vec{s}_i\cdot\vec{s}_j$ is smaller than $-Nz|A_1+A_2|$ obtained from the exchange Hamilton

$-2A \sum_{i<j} \vec{s}_i \cdot \vec{s}_j$ in the previous magnetism theorys. Formulas (13) and (14) demonstrate that the state of the system when $A_1 \neq |A_2|$ is the superposition of states $|1\rangle$ and $(|1\rangle+|2\rangle)$ or $|2\rangle$ and $(|1\rangle+|2\rangle)$, Weiss ferromagnetic state $|1\rangle$ or Neel anti ferromagnetic state $|2\rangle$ just a special state when $A_1=0$ or $A_2=0$.

## V.  SUMMARY

For the electrons system with the Heisenberg exchange integral $A = A_1 + A_2$ ($A_1>0$, $A_2<0$), only considering the overall exchange Hamiltonian as in the previous magnetism theories is unjustifiable, even wrong. For this electrons system, the correct exchange Hamilton is $H_{ex} = -2A_1 \sum_{i<j} \vec{s}_i \cdot \vec{s}_j - 2A_2 \sum_{i<j} \vec{s}_i \cdot \vec{s}_j$, not $-2A \sum_{i<j} \vec{s}_i \cdot \vec{s}_j$. The role of the minor term in the exchange Hamilton $H_{ex} = -2A_1 \sum_{i<j} \vec{s}_i \cdot \vec{s}_j - 2A_2 \sum_{i<j} \vec{s}_i \cdot \vec{s}_j$ must not be ignored. So the eigenstate of the system is

$$|X\rangle = \frac{1}{\sqrt{A_1^2 + A_2^2}}[(A_1 - |A_2|)|1\rangle + |A_2|(|1\rangle+|2\rangle)] \text{ or } |X\rangle = \frac{1}{\sqrt{A_1^2 + A_2^2}}[(|A_2|-A_1)|2\rangle + A_1(|1\rangle+|2\rangle)],$$

the energy eigenvalue is $E = -Nz(A_1 - |A_2|) - \frac{2Nz|A_2|}{A_1 + |A_2|} = -Nz(|A_2|-A_1) - \frac{2NzA_1}{A_1+|A_2|}$; when $A_2=0$, $|X\rangle=|1\rangle$, $E = -NzA_1$, the system is in Weiss ferromagnetic state (FM); when $A_1=0$, $|X\rangle=|2\rangle$, $E = -Nz|A_2|$, the system is in the Neel antiferromagnetic state (AFM); when $A_1 = |A_2|$, $|X\rangle = \frac{1}{\sqrt{2}}(|1\rangle+|2\rangle)$, $E = -NzA_1$, the probabilities of parallel and antiparallel arrange for every pair of spins in the system are equal, the system is in the spin glass state (SG); when $A_1 \neq |A_2|$, the probabilities of parallel and antiparallel arrange for every pair of spins in the system are not equal, the every pair of spins in the system is partly in the state $|1\rangle$ and partly in the state $|2\rangle$, and there coexist FM and SG or of AFM and SG for the whole system. Weiss ferromagnetic state or Neel anti ferromagnetic state just a special state when $A_1=0$ or $A_2=0$. The disorder orientation of spins in the SG state is induced by the competition between $A_1$ and $A_2$. The energy eigenvalue obtained from the exchange Hamilton $-2A_1 \sum_{i<j} \vec{s}_i \cdot \vec{s}_j - 2A_2 \sum_{i<j} \vec{s}_i \cdot \vec{s}_j$ is smaller than that obtained from the exchange Hamilton $-2A \sum_{i<j} \vec{s}_i \cdot \vec{s}_j$.

References

[1] D.C. Mattis, *The Theory of Magnetism II,* Springer Series in Solid-State Sciences, Vol. 17, ed. by Peter Fulde, (Berlin, Springer Verlag, 1985)

[2] S.V. Vonsovskii, *Magnetism,* Vol. 2, Translated from Russian by Rou Harclin, (New York, Halsted Press, 1974)

[3] T. Oquchi, *A Theory of Antiferromagnetism, II,* Prog. Theor. Phys. **13**, 148-159 (1955)

[4] J.A. Mydosh, *Spin Glasses: An Experimental Introduction*, (London, Taylor and Francis Inc., 1993)

[5] R. Blschof, E. Kaldis and P. Wachter, J. Magn. Magn. Mater., **31-34**, 255 (1983)

[6] F. Fujii, F. Pourarian and W.E. Wallace, J. Magn. Magn. Mater.,**27,** 215 (1982)

[7] Heisenberg W Z. Phsik 49, 619 (1928)

[8] KRITYUNJAI PRASAD SINHA AND NARENDRA KUMAR, Interaction in magnetically ordered solids (London, oxford university press,1980)

[9] Allan H. Morrish, The Physical Principles Of Magnetism (New York, IEEE Press, 2001)p.275

[10] Yongzhong Wang, Zhidong Zhang, solid state communications, 124, 215(2002)

[11] P. A . M. Dirac, Principle of Quantum mechanics (London, Oxford University Press, 1958)

[12] J. Y. Zeng，Quantum (in Chinese)（Beijing，Science Press，2000）Volume II , p.37